# Open source based cadastral information system: ANCFCC-MOROCCO

Hicham ELASRI[1], Mehdi NEKNANE[1], Jamila AATAB[1] and karima GANOUN[1]

[1] Department of Information System and Procedures (DISP), Rabat, Morocco

The National Agency of Land Conservation, Land Registry and Mapping

{h.elasri, me.neknane, j.aatab, k.ganoun}@ancfcc.gov.ma

**Abstract.** This present project is developing a geographic information system to support the cadastral business. This system based on open source solutions which developed within the National Agency of Land Registry, Cadastre and Cartography (ANCFCC) enabling monitoring and analysis of cadastral procedures as well as offering consumable services by other information systems: consultation and querying spatial data. The project will also assist the various user profiles in the completion of production tasks and the possibility to eliminate the deficiencies identified to ensure an optimum level of productivity.

## 1. Introduction.

Companies have a crucial need for agility. They must respond quickly to their business needs and suggest innovative solutions, also provide flexible and sustainable information system while facilitating the integration of new software components and applications. In this context, the coverage of Open Source solutions extends interest and convinces companies. The difficult economic environment companies require them to optimize costs and build value.

Faced with these restrictions, suppliers are trying to preserve their market share by influencing their customers, for example through audits or stricter licensing. This why it becomes increasingly difficult to negotiate with suppliers which will influence negatively on the customer-supplier relationship. Thus large firms naturally turn to the open source solutions because it seems more economical [1]. No doubt the open source optimizes the cost, flexibility and integration of new bricks unrestricted license. But the big worry is the risk that may be incurred at the lack of support or maintenance contract. Given all these factors, ANCCFC took the challenge of adopting an open source solution called gvSIG [1] to meet these business needs and also to overcome the limitations of the existing. The business that will be the subject of this work is the cadastre which interacted with other business: cartography and land register of ANCFCC. Our paper is organized as follow: First the problematic is presented. In section 3 the proposed solution. Finally, section 4 presents the conclusion and perspectives of our work.

## 2. Problematic

We developed a cadastral information system to support business activities of the cadastre and provide services to other business in ANCFCC. The FIG statement on Cadastre (FIG, 1995) defines cadastre as follows: « *A cadastre is normally a parcel based, and up-to-date land information system containing a record of interests in land (e.g. rights, restrictions and responsibilities). It usually includes a geometric description of land parcels linked to other records describing the nature of the interests, the ownership or control of those interests, and often the value of the parcel and its*

---

[1] http://www.gvsig.org/web/

*improvements. It may be established for fiscal purposes, legal purposes, or to assist in the management of land and land use and enables sustainable development and environmental protection.*[2].

There is a strong relationship between the cadastre, land registration and cartography. Although the Land Registry holds the records for land rights through acts or title, cadastre contains technical information on properties and their limitations in an administrative domain, and the cartography contains geodetic information. Activities resulting from trades of land registration, cartography and cadastre are mutually complementary and should ideally be treated in the same system. The second statement of Cadastre model 2014 [3] provides an abolition of the separation between cadastral, cartography and land registers. Yet, in many cases, they work independently in separate organizations and there is not always joint operation in the most effective way [4]. In our case the cadastral information systems, cartography and land are autonomous with some interoperability problem.

ANCCFC currently has a geographic information system called MAPINFO which is an extension CADGIS for the management of cadastral data. MAPINFO or specifically purchased by ANCFCC version has several limitations:

- Data is stored in flat files which cause a difficulty for research, archiving, security and the big problem is the lack of multi-user access.
- Proprietary solution, difficulty to add others extensions.
- Proprietary file format (DAT, MAP, TAB…).
- Only graphic data are managed by CADGIS.
- There is no direct link between the graphical data and alphanumeric data.

Difficult interoperability with existing tools either in syntactic or semantic level. This is due to several types of heterogeneity: syntactic, semantic and structural.

For example the pivot between the different data sources ANCFCC is the Title number that is written in different formats.

|  | Format 1 | Format 2 | Format 3 |
|---|---|---|---|
| Title number | **NatureNumber/index** (example T1111/20) | **Nature/Number/index** (example T/1111/20) | **Number conservation land / Nature/Number/index** (example 03/T/1111/20) |

Table 1. Syntactic heterogeneity of the title.

The formats of the title number represents a syntactic heterogeneity, the well-known in the literature to answer this type of heterogeneity is the conversion approach. In other types: semantic and Structural heterogeneity, we need a kind of solution proposed by [5].

## 3. The proposed solution

In order to overcome the limitations of existing we developed a cadastral information system called GEO-SIC based on the open source gvSIG and PostgreSQL / PostGIS (Our choice of gvSIG is derived from the comparative study of open sources proposed by [2]). By implementing the approach presented in Figure 1.

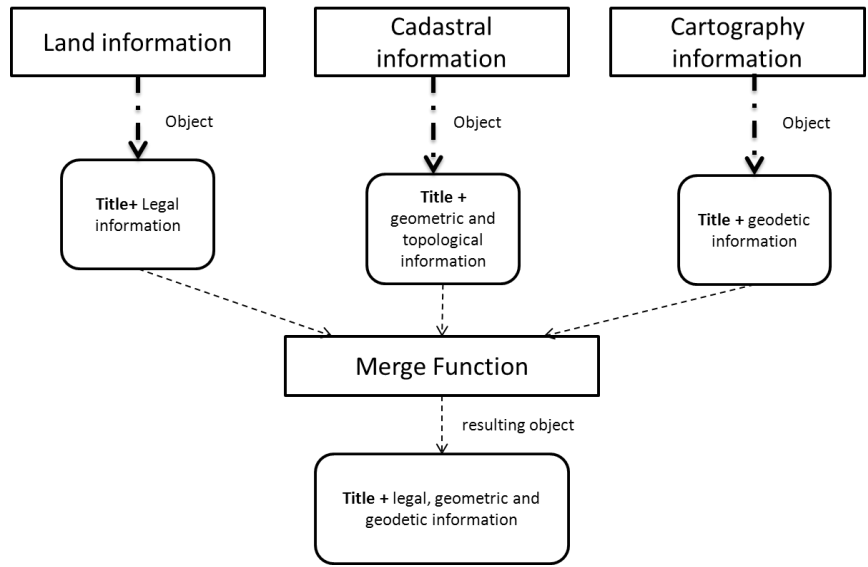

Fig. 1 Merging objects from different sources.

GEO-SIC is a Geographic Information System (GIS) which provides not only management information but also graphics and alphanumeric data documentaries to solve complex problems of management and production data. It has a wide range of features to process geographic information. Among its features there is the possibility of geographic consultation, attributing data, data storage, geographic data updating, spatial analysis, data creation and editing, georeferencing raster files, reprojection Management, Geoprocessing (Buffer, intersection, union, spatial joins, selection ...), Realization of treatment on vector data and thematic analysis performance. All this tools are open source and multi-systems (Windows, MacOS X, and Linux). So it can work on remote servers like MySQL, PostGIS and Oracle (with the extension GeoDB). It provides multiple versions of multiple contexts: mobile, desktop. But, actually it doesn't offer a web version and no connectors for SQL SERVER SPATIAL 2008.

This information system has gone through several phases of which two are essential: development phase and migration data.

In the **development phase** we have developed components for the management, treatment of cadastral data in a first step and we will develop components to meet business requirements mapping and land registration in a second step. Figure 2 shows a component added to the gvSIG which supports a cadastral activity.

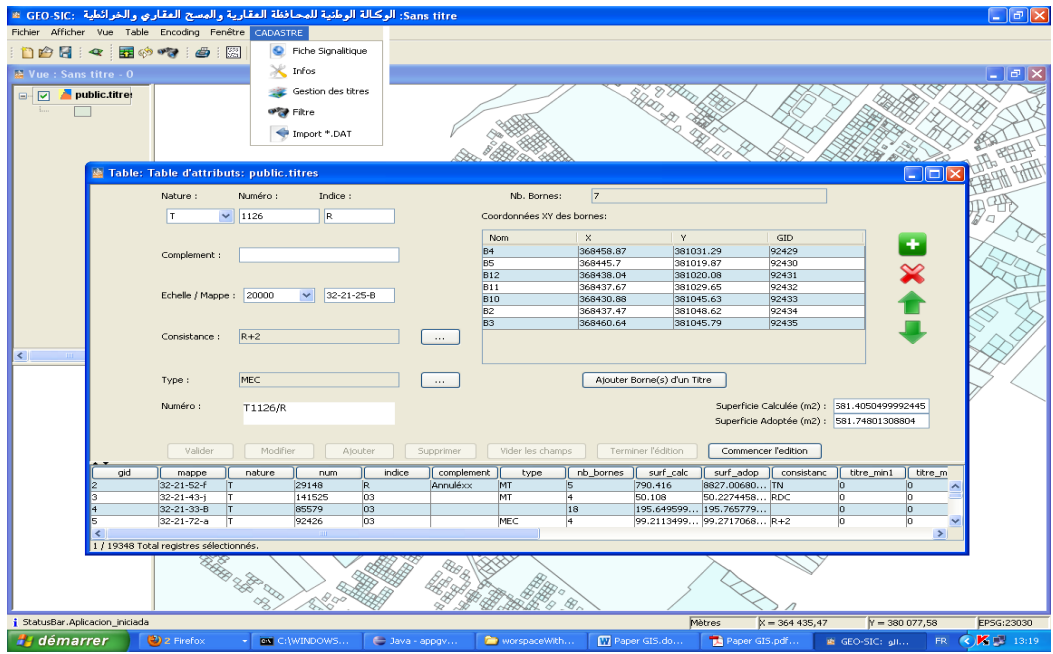

Fig. 2. GvSIG extension "CADASTRE" to give the management of cadastral

**Migration phase:** This phase involves the data migration of MAPINFO/CADGIS to a PostGIS database (Figure 3).

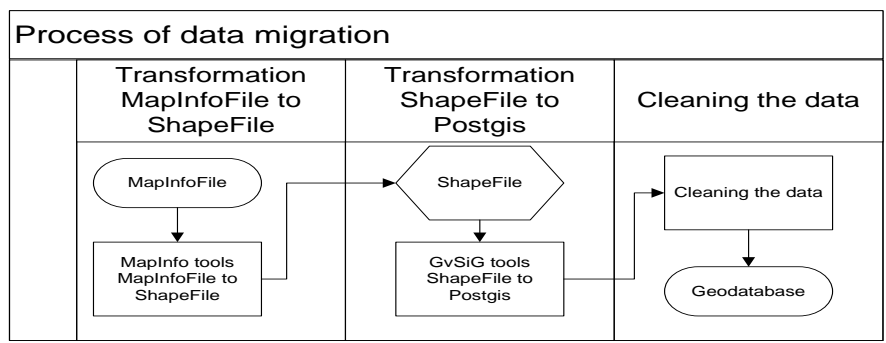

Fig. 3. Migration process of cadastral data.

## 4. Conclusions.

We have developed several extensions for the management of cadastral data: map management, management titles ... Currently we are developing a deployment strategy for the whole territory of Morocco. We focus to develop or use a web version for publishing geographic data. In order to optimize our architecture, we focus in future work to define interoperability models and see which approach measures the service oriented architecture (SOA) can help in our Problematic.

# Open source based cadastral information system: ANCFCC-MOROCCO

Hicham ELASRI[1], Mehdi NEKNANE[1], Jamila AATAB[1] and karima GANOUN[1]

[1] Department of Information System and Procedures (DISP), Rabat, Morocco

The National Agency of Land Conservation, Land Registry and Mapping

{h.elasri, me.neknane, j.aatab, k.ganoun}@ancfcc.gov.ma

*Keywords:* GIS, gvSIG, Cadastre.

Le theme: *Spatial Data Infrastructure*